\documentclass[pra,aps,amsfonts,amsmath,superscriptaddress,twocolumn,showpacs,floatfix]{revtex4}
\usepackage{epsfig}

\begin{document}

\title{Suppression, persistence and reentrance of superfluidity in overflowing nuclear systems}

\author{J. Margueron}
\author{E. Khan}
\affiliation{Institut de Physique Nucl\'eaire, Universit\'e Paris-Sud, IN2P3-CNRS, F-91406 Orsay Cedex, France}

\begin{abstract}
Based on a microscopic description of superfluidity in overflowing nuclear systems, 
it is shown that continuum coupling plays an important role in the suppression, the
persistence and the reentrance of pairing.
In such systems, the structure of the drip-line nucleus determines the suppression and the persistence of superfluidity.
The reentrance of pairing with increasing temperature leads to additional
critical temperatures between the normal and superfluid phases.
\end{abstract}

\pacs{67.85.Lm, 74.40.Kb, 21.60.Jz, 26.60.Gj}

\date{\today}

\maketitle

Overflowing many-body fermionic systems exist in various situations going from the crust of neutron 
stars~\cite{Book:Haensel2007} to ultra-cold atoms~\cite{Stamper-Kurn1998,Viverit2001}.
Interestingly, these systems offer the possibility to study the coupling between two fluids with very 
different pairing properties~\cite{Grasso2008,Schuck2011}.
In such systems, one fluid is localized inside an initial container, such as for instance a nuclear potential, 
and a second fluid is overflowing towards a larger container.
Being in different environments, these two fluids can acquire different pairing gaps.
In this Letter, we address the question of the coupling between the two superfluids and their finite temperature
properties.

At the transition between its outer and inner crust, neutron stars provide an example of such
microscopic overflowing systems, commonly called neutron dripping~\cite{Book:Haensel2007}.
In the outer crust, nuclei form a Coulomb lattice which gets more and more neutron rich as the density increases.
When the maximum number of neutrons that a nucleus can sustain is reached, the 
overproduced neutrons drip out of nuclei.
These neutrons populate the continuum states 
and shall be described within the band theory (see Refs.~\cite{Book:Ashcroft1976,Chamel2007} and 
references therein).
It should be noted that nuclei surrounded by an infinite neutron gas could exist as a stable 
configuration in neutron stars where they are bounded by gravitation, while isolated nuclei that exist for 
instance on earth are limited to the drip lines.

The first prediction of the suppression of pairing in overflowing $Z=50$ nuclear systems was performed in 
Ref.~\cite{Grasso2008}.
It was proposed to attribute this suppression to the large coherence length 
of the weakly-superfluid neutrons gas: the neutron gas can penetrate the dense nuclear system and
could impose its weak pairing field.
It was however also noted in the same work that neutrons are dripping out of the double-magic 
nucleus $^{176}$Sn. 
In a recent work, the pairing gap of the last occupied state was also predicted to be quenched in overflowing 
nuclear systems, metallic grains and cold-atoms~\cite{Schuck2011}.
It was concluded that the suppression of superfluidity is a generic fact of a fermionic superfluid
overflowing from a narrow container into a much wider one.
It shall be noted that in this work the nuclear model was mainly limited to $\ell$=0 single particle states
(s-states).
The description of nuclear systems requires however the inclusion of $\ell > 0$ single particle states 
for the bound and the mean-field resonance states.
It is indeed well known that, close to the drip lines, resonance states play an important role for 
pairing properties and it is referred generally as continuum coupling~\cite{Bulgac1980,Grasso2001}.
It has recently regained some attention in nuclei close to the drip-line (see for instance 
Refs.~\cite{Zhang2011,Hagino2011} and references therein).
The general question of continuum coupling in overflowing superfluid systems and various phenomenons 
such as persistence, suppression and reentrance of pairing remains to be studied.
Pairing correlations in the ground state of weakly-bound nuclei are commonly described by the 
Hartree-Fock-Bogoliubov (HFB) theory~\cite{Book:DeGennes1989,Book:Ring1980}.
In most of the HFB calculations the continuum is discretized by solving the HFB equations with box 
boundary conditions~\cite{Dobaczewski1984a}.
In this Letter a systematic analysis based on several overflowing isotopic chains is performed and it is shown 
that pairing quenching and continuum coupling are strongly related.
A pairing reentrance phenomenon with increasing temperature is predicted for $Z=50$ overflowing systems.


In the present work, an HFB approach in coordinate representation is employed.
This model has already been applied to describe nuclei and Wigner-Seitz cells in a fully self-consistent 
framework (see Ref.~\cite{Grill2011} and references therein).
The Skyrme SLy4 interaction~\cite{Chabanat1998a} is used in the mean-field channel, and is completed 
by the ISS pairing force which is adjusted to the BCS pairing gap predicted by bare nucleon-nucleon 
potentials~\cite{Grill2011}.
All the bound states are considered, the angular momentum goes up to $J_{max}=27/2$, and 
we take a large box radius $R_{box}$=26.7~fm ensuring an adequate description of the continuum.
An almost constant neutron density at the edge of the Wigner-Seitz cell is obtained using mixed 
Dirichlet-Von Neumann boundary conditions.
The expected nuclei in neutron stars located at the transition between the outer crust and the
inner crust have a proton number Z around 30 to 50, depending on the models~\cite{Ruster2006}.
We therefore selected 8 isotopes given in Table~\ref{L11:tab:drip}  and located around Z=28, 40 and 50.
It is well adapted to perform a spherical HFB calculation since these nuclei are predicted spherical 
near the neutron drip-line~\cite{Hilaire2007}.

\begin{table}[t]
\setlength{\tabcolsep}{.08in}
\renewcommand{\arraystretch}{1.4}
  \begin{center}
  \begin{tabular}{ccccc}
    \toprule
    Isotope & $Z$ & $N_{drip}$ & group & $N_{res}$ \\
    \colrule
    Ni & 28 & 60 & $\mathcal{A}_1$ & 3.0 \\
    Kr & 36 & 82 & $\mathcal{A}_2$ & 0.0 \\
    Sr & 38 & 82 & $\mathcal{A}_2$ & 0.0 \\
    Zr & 40 & 84 & $\mathcal{A}_1$ & 2.2 \\
    Mo & 42 & 90 & $\mathcal{A}_1$& 8.0 \\
    Ru & 44 & 92 & $\mathcal{A}_1$& 3.0 \\
    Sn & 50 & 126 & $\mathcal{A}_2$& 0.0\\
    Te & 52 & 126 & $\mathcal{A}_2$ & 0.0\\
    \botrule
  \end{tabular}
  \end{center}
  \caption{Isotope acronym, number of protons $Z$, number of neutrons of the last nucleus before the drip line $N_{drip}$, 
  for the selected set. 
 Isotopes for which the drip-line nucleus is non-magic (magic) belongs to the group $\mathcal{A}_1$ 
 ($\mathcal{A}_2$, respectively). 
 The total occupation number of resonance states for the drip-line nuclei $N_{res}$ is shown in the last 
 column (see the text for more details).}
  \label{L11:tab:drip}
\end{table}

For these selected isotopes, the neutron-drip number $N_{drip}$, defined as the neutron number of the 
last nucleus before the two-neutrons separation energy $S_{2n}(N)=E(N)-E(N-1)$ changes its 
sign~\cite{Book:Ring1980}, is given in table~\ref{L11:tab:drip}.
The total neutron occupation number of resonance states $N_{res}$ for the drip-line nuclei is
also given.
The calculation of $N_{res}$ requires the identification of resonance states which, in our case, 
are defined as positive energy states with rms radii lower than 10~fm.
It is interesting to remarck from Table~\ref{L11:tab:drip} that the drip-line nuclei with a number of
neutrons that coincides with the well-known magic numbers (82 and 126) have no states in the
continuum. 
The shell occupation in this case is 0 or 1, as in magic nuclei.
We therefore group Kr, Sr, Sn and Te isotopes in the same group $\mathcal{A}_2$.
The persistence of magicity with the same magic numbers as in stable nuclei might be due to
the sphericity of these nuclei.
In the case of Ni, Zr, Mo and Ru, the drip-line occurs in more complicated shell structure 
partially involving resonance states and these isotopes are grouped in $\mathcal{A}_1$.

The drip-line isotopes identified in Table~\ref{L11:tab:drip} can be considered as the seed nuclei from which the 
overflow occurs.
The dripping of neutrons shall therefore be influenced by the microscopic structure of these seed nuclei.
Nuclei belonging to the group $\mathcal{A}_1$ have a resonance occupation number $N_{res}$ which goes
from 2 to 8 particles. This number reveals the large continuum coupling.
The number of particles in the resonance states is zero for nuclei belonging to the group $\mathcal{A}_2$, since in 
the magic nuclei located at the drip-line, pairing correlations are quenched and continuum coupling therefore 
largely suppressed.
From a microscopic analysis, the structure of drip-line nuclei is qualitatively different between the group 
$\mathcal{A}_1$ and $\mathcal{A}_2$.
In the following, we show how these differences impact the quenching of pairing correlations at overflowing.

\begin{figure}[t]
\begin{center}
\includegraphics[width=0.95\linewidth]{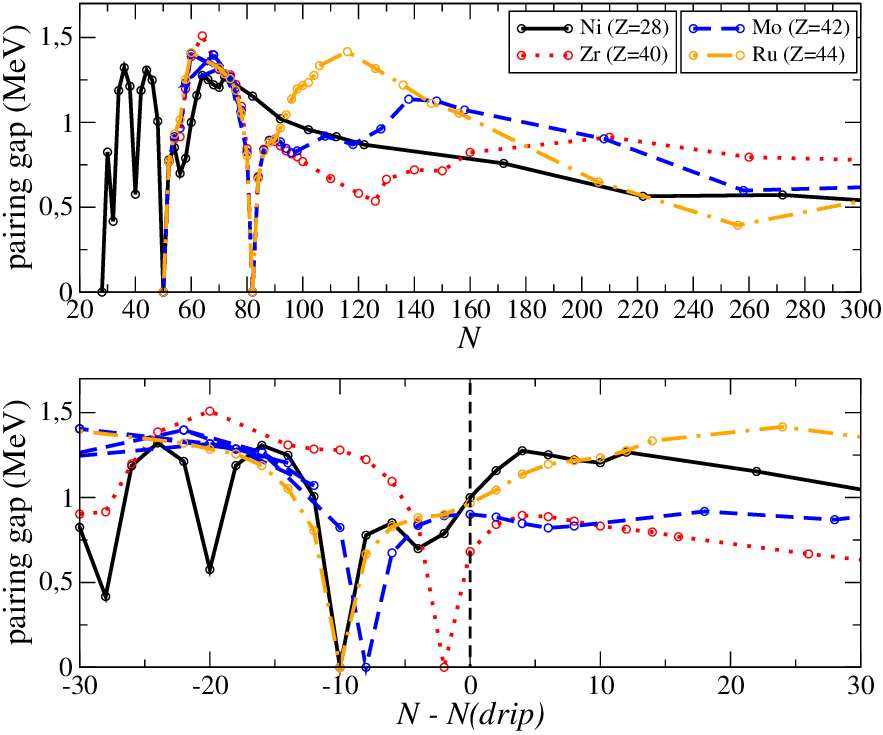}
\caption{(color online) Neutron pairing gaps versus neutron density $N$ (top panel) and versus
$N-N(drip)$ (bottom panel) for the isotopes in the group $\mathcal{A}_1$ (see the text for more details).}
\label{L11:fig:fig02}
\end{center}
\end{figure}

\begin{figure}[t]
\begin{center}
\includegraphics[width=0.95\linewidth]{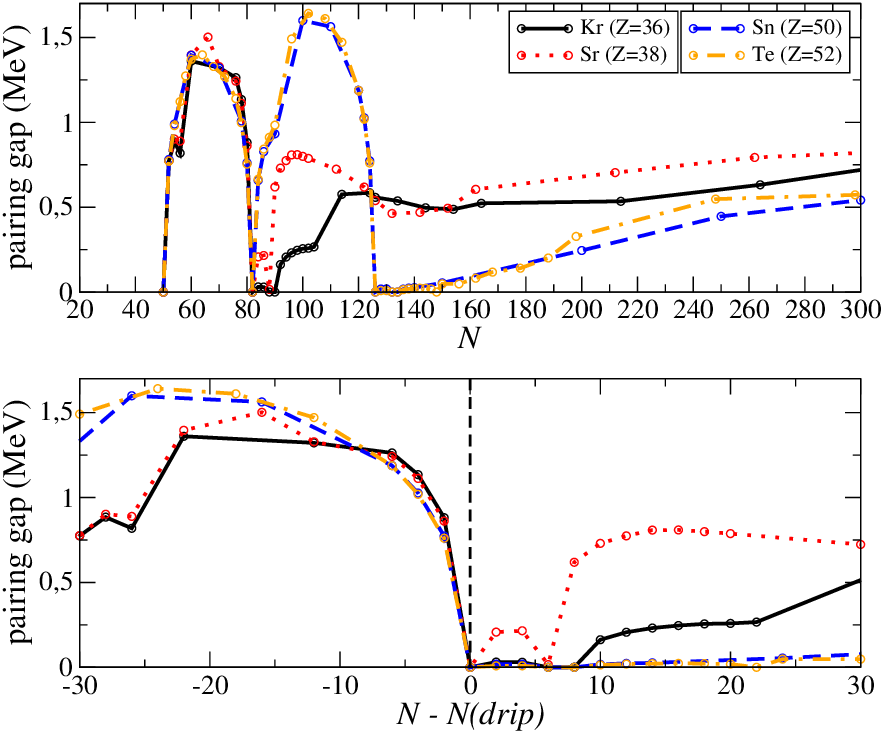}
\caption{(color online) Same as Fig.~\ref{L11:fig:fig02} for isotopes in the $\mathcal{A}_2$ group. }
\label{L11:fig:fig03}
\end{center}
\end{figure}

Figs.~\ref{L11:fig:fig02} and \ref{L11:fig:fig03} display the average neutron pairing gaps versus the 
neutron number (top panel) and the difference between the neutron number and the neutron drip 
$N-N_{drip}$ (bottom panel).
The average neutron pairing gap is obtained from the neutron pairing field given by the HFB  
solution.
Fig.~\ref{L11:fig:fig02} represents the neutron pairing gaps for the $\mathcal{A}_1$ isotopes while in 
Fig.~\ref{L11:fig:fig03} are shown the neutron pairing gaps for the $\mathcal{A}_2$ isotopes.
In the case of nuclei from the $\mathcal{A}_1$ group, the neutron pairing gap persists beyond the
drip line (see bottom panel of Fig.~\ref{L11:fig:fig02}).
The continuum coupling preserves the pairing diffusivity around the Fermi energy: the occupancy
of the scattering states beyond the drip-line does not suppress superfuidity.
The presence of an overflowing neutron gas has therefore a limited effect on the pairing gap 
for the $\mathcal{A}_1$ isotopes, showing the persistence of superfluidity.
In the case of the $\mathcal{A}_2$ isotopes, a suppression of the average pairing gap just beyond 
the drip line is observed (Fig.~\ref{L11:fig:fig03}).
The presence of a magic nucleus at the drip line have therefore a strong influence on the pairing gap 
at and beyond the drip line: the large spacing between the last occupied state and the first excited state
(resonance state) suppresses superfluidity.
Figs.~\ref{L11:fig:fig02} and \ref{L11:fig:fig03} provide a clear illustration that the shell structure of the 
drip-line nucleus determines the suppression or the persistence of superfluidity, even in overflowing
nuclear systems.

In Ref.~\cite{Schuck2011}, the quenching of pairing in overflowing systems has been predicted 
upon overflow of trapped fermions.
It should be underlined that in the present work, the mean-field in nuclear systems have a centrifugal term 
which gives rise to the mean-field resonance states and continuum coupling,
whereas the conclusions obtained in Ref.~\cite{Schuck2011} are applicable only to overflowing systems involving
$\ell=0$ single particle states.
In more general systems, like in the crust of neutron stars, a more sophisticated theoretical 
model such as the present HFB one predicts a suppression of superfluidity only if the continuum 
coupling is quenched, by the shell structure for instance.

\begin{figure}[t]
\begin{center}
\includegraphics[width=0.95\linewidth]{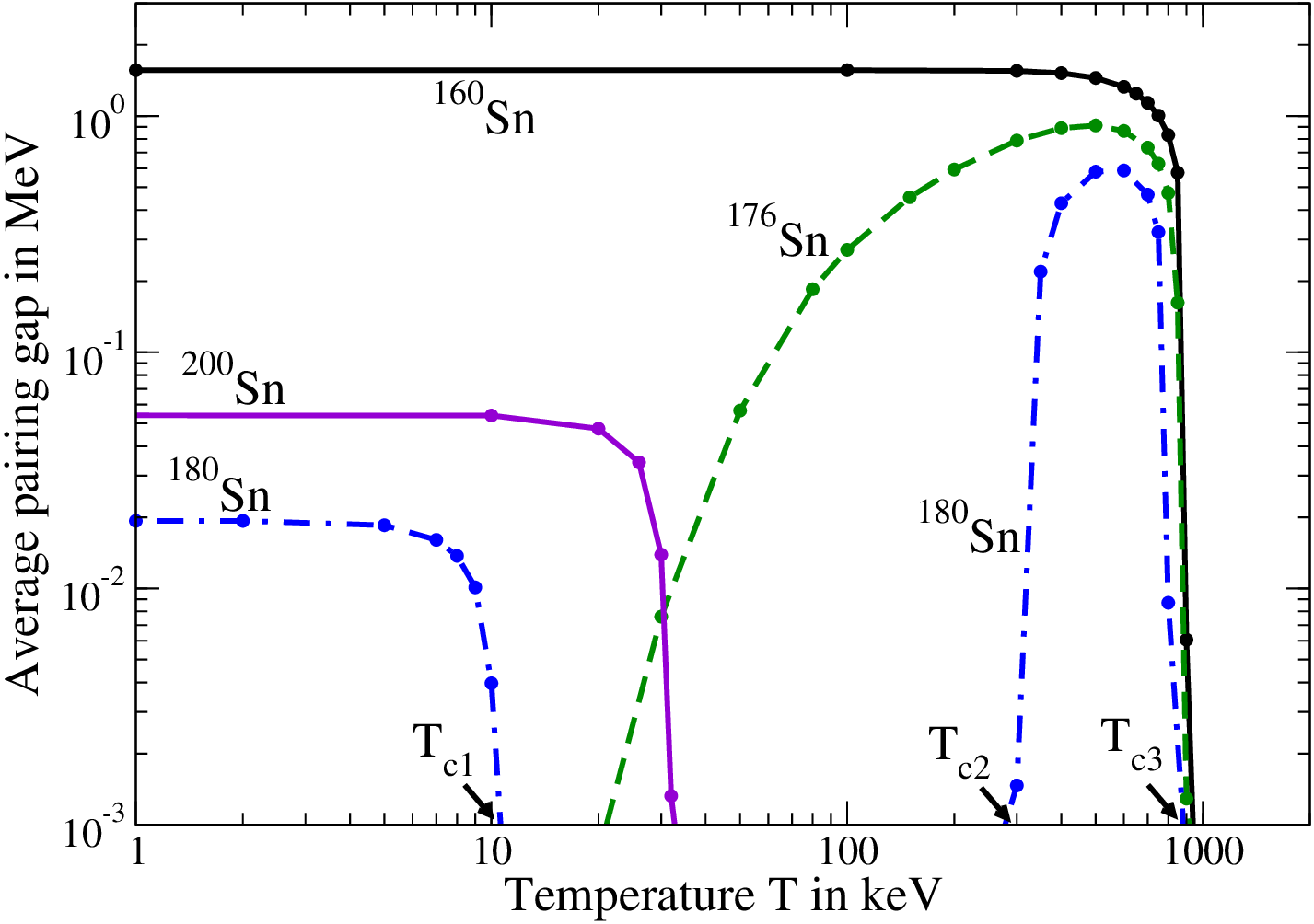}
\caption{(Color online) Temperature averaged neutron pairing gap versus temperature for
$^{160, 176, 180, 200}$Sn.
A superfluid re-entrant effect is observed for $^{176}$Sn and $^{180}$Sn.}
\label{L11:fig:fig3c}
\end{center}
\end{figure}

The persistence of superfluidity upon overflow of bound neutrons might not be the only consequence of
continuum coupling: in the case where pairing is suppressed, the increase of temperature may generate
the reentrance of superfluidity.
The finite-temperature HFB model~\cite{Sandulescu2004},
is employed to study the reentrance of pairing in the thermal state of overflowing 
even-even nuclear systems.
The temperature-averaged pairing gap for $^{160,176,180,200}$Sn is shown in Fig.~\ref{L11:fig:fig3c}.
In $^{160}$Sn and $^{200}$Sn it behaves as expected from 
HFB theory: the pairing gap vanishes at the critical temperature $T_c=0.57 \Delta(T\approx 0)$  (see 
Ref.~\cite{Khan2007} and references therein).
In the case of $^{176}$Sn and $^{180}$Sn, the reentrance of superfluidity is observed with increasing
temperature.
This reentrance is induced by the presence of resonances states in the spectrum of these nuclear systems: 
Being slightly too high in energy, these states are not occupied at zero temperature (see 
Table~\ref{L11:tab:drip}), 
while at finite temperature, they can be
partially occupied 
from the Fermi-Dirac distribution.
At low temperature, the pairing correlations can therefore be switched on allowing the reapparition of the
superfluid state.
The reentrance critical temperature depends on the step in energy between the last occupied bound state and the
first resonance one, which changes from one system to another, as observed in Fig.~\ref{L11:fig:fig3c} for $^{176}$Sn 
and $^{180}$Sn.
In the case of $^{200}$Sn, this energy step is too large to give rise to the reentrance of superfluidity before 
the highest critical temperature is reached.
The $^{180}$Sn overflowing system has an interesting phase diagram including three critical temperatures:
with increasing temperature, two of them correspond to the vanishing of superfluidity ($T_{c1}\sim11$~keV and 
$T_{c3}\sim1$~MeV) and one to its reappearance ($T_{c2}\sim300$~keV).
The lowest critical temperature $T_{c1}$ is associated to the transition from the superfluid to the normal state 
in the overflowing neutron gas.
The highest critical temperature $T_{c3}$ is similar for $^{160,176,180}$Sn, indicating that superfluidity has 
been restored in the seed nucleus of $^{176,180}$Sn between $T_{c2}$ and $T_{c3}$. 
This superfluidity is mainly built on resonances populated at finite temperature.
More generally, pairing reentrance in hot systems is observed for isotopes belonging to the group 
$\mathcal{A}_2$ where resonances are too high in energy to participate to pairing at zero temperature but close
enough to the last occupied state to be reached at finite temperature.

Reentrance of superfluidity at finite temperature have been predicted in nuclear systems such as 
in odd-nuclei~\cite{Balian1999}, rotational motion of nuclei~\cite{Dean2010}, and the deuteron pairing
channel in asymmetric infinite matter~\cite{Sedrakian1997}.
It was aslo predicted in polarised $^{3,4}$He~\cite{Frossati1986,Crowell1993,Csathy2002} and in 
spin asymmetric cold atom gas~\cite{Castorina2005,Levin2006}.
In all these systems, pairing at zero temperature is generated by an attraction among Fermions of different
spin or isospin.
Superfluidity is therefore maximum in spin or isospin symmetric systems for which there is a matching of 
the Fermi levels of the constituent Cooper-pairs.
Breaking the spin or isospin symmetry disfavor pairing while temperature in asymmetric systems acts 
in favor of restoring the broken symmetry and can eventually induce a reentrance of pairing.
At variance with this mechanism the pairing reentrance phenomenon discussed in this Letter is based on 
a novel mechanism in finite systems where resonance states play a major role.


In summary, we have investigated the pairing properties of nuclear systems upon overflowing superfluid 
neutrons.
Suppression, persistence and reentrance of superfluidity can occur in these finite systems.
From a systematic HFB calculations on 8 isotopic chains, the pairing properties is shown to be 
strongly correlated to the continuum coupling, both at zero and finite temperature.
At zero temperature, the coupling between the seed nucleus and the gas is weak, and a formal 
separation of their properties into a nucleus plus a gas provides a qualitative understanding of the 
suppression and the persistence of superfluidity.
With increasing temperature in the normal state, the Fermi-Dirac distribution can populate the resonance 
states giving rise to the reentrance of superfluidity.
The pairing correlations in the nuclear system are switched on again and consecutive critical temperatures
are predicted.

The understanding of the suppression, persistence and reentrance of superfluidity in nuclear systems, 
deeply related to the continuum coupling, opens wide perspectives for discoveries in weakly bound 
nuclei, as well as it sheds new light 
on the transition between the inner and outer crusts in neutron stars.
The role of resonances around the neutron drip not only changes the microscopic understanding of the 
neutron drip-out mechanism, but it also modifies the thermal properties of the crust through the strength of the
pairing interaction~\cite{Fortin2010}.
The temperatures at work during cooling are typically of the order of 10 to 
500~keV~\cite{Book:Haensel2007} 
and coincide with the critical temperatures of the reentrance phenomenon.
As a consequence, the novel pairing reentrance phenomenon analyzed in this Letter is expected to
modify the thermodynamical and cooling properties in the crust of neutron stars.
The links with other superfluid Fermi-systems shall be investigated in the near future.
For instance in cold Fermionic atoms overflowing from an inner trap to a larger
one~\cite{Stamper-Kurn1998,Viverit2001}, it is known that stable path close to the centrifugal energy
are the classical analog of the quantal resonances.
It will therefore be interesting to investigate the role of these stable paths on the superfluid 
properties of overflowing cold Fermionic atoms.

The authors thanks M. Baldo, P. Schuck, X. Vi\~nas and M. Urban for the interesting discussions during the completion 
of this work.
This project was partially supported by the ANR NExEN and SN2NS contracts, the Institut Universitaire de France and 
by COMPSTAR, an ESF Research Networking Program.


\end{document}